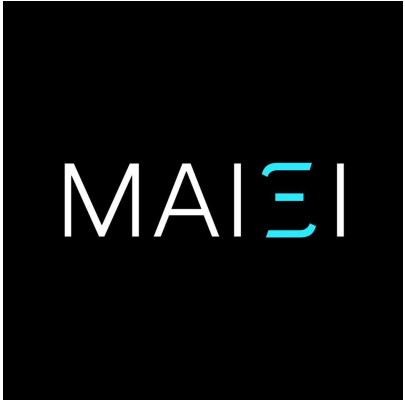

# Montreal AI Ethics Institute

*An international, non-profit research institute helping humanity define its place in a world increasingly driven and characterized by algorithms*

**Website**: https://montrealethics.ai
**Newsletter**: https://aiethics.substack.com

# Report prepared by the Montreal AI Ethics Institute for the Santa Clara Principles for Content Moderation

*Based on insights and analysis by the Montreal AI Ethics Institute (MAIEI) staff on the Santa Clara Principes and supplemented by workshop contributions from the AI Ethics community convened by MAIEI on June 17th and June 25th*

Primary contacts for the report:

Marianna B. Ganapini, Researcher
marianna@montrealethics.ai

Camylle Lanteigne, Researcher
camylle@montrealethics.ai

Abhishek Gupta, Founder
abhishek@montrealethics.ai

**Overview of our recommendations:**

- There should be more diversity in the content moderation process. Potential biases and discriminatory decisions constitute a great concern for content moderation whether performed by a human or machine.
- There is a need for transparency concerning how platforms guide content-ranking, which has the potential to restrict freedom of expression and users' autonomy, and stifle social change.
- Anonymized data on the training and/or cultural background of the content moderators employed by a platform should be disclosed.
- There are no one-size-fits-all solutions: content moderation tools must be tailored to specific issues. For instance, misinformation may be best addressed through behavioral nudges, whereas hate speech may require more drastic measures. Guidelines to address *all* the possible types of content moderation tools employed on online platforms are necessary.
- Specific guidelines are needed for messaging applications with regards to data protection in content moderation.
- Cultural differences relevant to what constitutes acceptable behavior online need to be taken into account in content moderation.
- When it comes to political advertising, we need to make sure that platforms are transparent. Integrity policies for political content should be the same as the policies adopted for other types of content.
- The flagging/reporting system provided to users by platforms would benefit from greater transparency, as it may be particularly problematic when used in contexts where the majority of users are prone to discriminate against minority groups
- When user content is flagged or reported, it must be clear when the flagging and reporting is automated.
- More data should be made available on the *types* of content removed from platforms online to make this process more transparent.
- Platforms should provide clear guidelines on the appeal process, as well as data on prior appeals. The appeal process should also be intelligible to a layperson, and not make one feel as though they must seek external legal counsel to navigate said process.
- We believe the Principles should be periodically revisited -- for instance, every two years -- or within a timeframe that allows for any appropriate revisions. This would allow the Principles to reflect various technological advancements, modifications in law and policy, as well as changing trends or movements in terms of platforms' content moderation.

**Introduction:**

In the United States, Section 230 of the Communications Decency Act gives online platforms vast immunity from liability for user-generated content appearing on their sites. This provision was designed to strike a balance between, on the one hand, pushing platforms to moderate their content and, on the other, allowing for freedom of expression.[1] The best practices around content moderation, however, are difficult to shape and formulate because the real nature of these online platforms has proven hard to determine: should these platforms be seen as publishers, editors or simple broadcasters of their content? And given this, what is the right way to ensure freedom of expression while also avoiding offensive and harmful online content?

As we are trying to figure out the answers, we know that online platforms are already heavily curating their content.[2] The Santa Clara Principles (SCP) are intended to offer some guidelines on how to make this process more transparent and ultimately more fair. We believe that the SCP are a very valuable tool and here we would like to offer some suggestions for improvement. In what follows, we go over each one of the questions in the call for submissions. We offer our recommendations based on insights and analysis by the Montreal AI Ethics Institute (MAIEI) staff and supplemented by workshop contributions from the AI Ethics community convened by MAIEI during two online public consultation meetups.

**1. Currently the Santa Clara Principles focus on the need for numbers, notice, and appeals around content moderation. This set of questions will address whether these categories should be expanded, fleshed out further, or revisited.**

> a. The first category sets the standard that companies should publish the numbers of posts removed and accounts permanently or temporarily suspended due to violations of their content guidelines. Please indicate any specific recommendations or components of this category that should be revisited or expanded.

We agree that it is essential to have a clear picture of the numbers of the posts flagged and removed as a result of content moderation. The more fine-grained data we can have the better it

---

[1] Social media platforms are private companies, and the First Amendment of the US Constitution does not directly apply to them. However, there are clear incentives for these platforms to both moderate their content *and* preserve freedom of expression. For an analysis and historical overview of these reasons see Klonick, K. (2017). *The New Governors: The People, Rules, and Processes Governing Online Speech* (SSRN Scholarly Paper ID 2937985). Social Science Research Network. https://papers.ssrn.com/abstract=2937985

[2] See, for instance: Weedon, J., Nuland, W., & Stamos, A. (2017). *Information Operations and Facebook*. Facebook. https://fbnewsroomus.files.wordpress.com/2017/04/facebook-and-information-operations-v1.pdf; *Community Guidelines*. (2020, January). TikTok. https://www.tiktok.com/community-guidelines?lang=en; and Buni, C. (2016, April 13). *The secret rules of the internet*. The Verge. https://www.theverge.com/2016/4/13/11387934/internet-moderator-history-youtube-facebook-reddit-censorship-free-speech

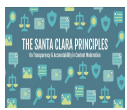

is to understand the overall impact of content moderation. In particular, we believe that it is important that platforms and companies disclose the "number of discrete posts and accounts flagged, and number of discrete posts removed and accounts suspended, by locations of flaggers and impacted users (where apparent)."
However, we believe that more fine-grained data is needed, at least concerning the following three areas:

Data on **users' flagging/reporting system**: one worry we have is that the users' flagging/reporting system -- that allows users to voice their concerns about content -- may be particularly problematic when used in contexts where the majority of users are prone to discriminate against minority groups and thus see their posts as 'offensive'.

To address these and other kinds of issues, platforms (e.g. Facebook) may not only ask users to categorize the type of harm they see in the posts they reported, but they may also push them to engage in "social reporting" especially if the flagged content does not violate any of the platforms' internal rules or community standards.

As a result, to obtain a more granular picture of the content moderation process, it is important that these platforms disclose the numbers of initially *flagged* posts and how many of them are:

(i) prioritized and reported to the authorities because they raised serious concerns based on internal rules or standards;
(ii) categorized as less urgent even if they are still violating internal rules or standards;
(iii) seen as *not* violating internal rules or standards even though they prompted "social reporting".

We believe that one of the goals of this numbers-disclosure is to possibly reveal patterns of discrimination against certain communities and to get a better sense of how users' reporting methods are used and interpreted by each platform.

**Data on topics of removed posts/accounts:** when accounts are removed it would be useful to have more data on the type of content or topics of their posts. For instance, Twitter revealed it actively blocks coordinated misinformation campaigns on the platform. Recently, they "removed more than 1,600 accounts originating in Iran. Cumulatively, these accounts Tweeted nearly 2 million times […] often with an angle that benefited the diplomatic and geostrategic views of the Iranian state."[3] Similarly, we ask for more data and transparency on the types of content removed from platforms online.

---

[3] Roth, Y. (2019, June 13). Information operations on Twitter: Principles, process, and disclosure. *Twitter Blog*. https://blog.twitter.com/en_us/topics/company/2019/information-ops-on-twitter.html

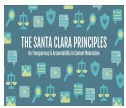

**Timing, number of appeals and reversed decisions**: platforms need to be explicit about the turnaround time between the moment something is reported/flagged and when it is taken down. They should also reveal the number of appeals and reversed decisions, and their locations.[4]

We also ask that in this reporting process as more of the flagging and reporting is being done by automated means[5], that there be a degree of transparency on which of the decisions were made by automated means, using what criteria, and how many of them were subsequently upheld by human reviewers where the confidence threshold for the system was low such that it was passed on to human reviewers. When such a thing happens, it would be desirable to see where the differences were between the decision made by the automated system and the human reviewer. For ones that were not reviewed by a human, a clear documentation of the criteria against which the decision was made and the confidence levels of the system should be recorded for audit purposes. Another example where automated systems can have inherent biases that might negatively affect a group are people with disabilities who face a disproportionate burden in terms of censorship when it comes to how they express themselves and the content that relates to or speaks about disabilities.[6]

In automated systems that are used for content flagging, if a platform is using an external service provider (e.g. Perspective API[7] from Google to detect the toxicity of content) we also advocate for the platform to make a transparent declaration of that. This serves the community in terms of pointing out potential concerns that arise from the work done by external researchers for problems in that tool / service. Additionally, requesting for limitations of such third-party services should be included in the content moderation policies as it will help to inform the decision that a user might make when it comes to whether or not to appeal the decision, especially when it might relate to known flaws in the system as it relates to specific types of content.

When looking at some of the quantitative measures proposed under the first principle, for larger public understanding, we believe it will be useful to also provide these numbers as percentages to give the lay public an understanding of the prevalence of content moderation which might otherwise not be apparent from raw numbers since the volume of content defies everyday understanding in terms of the orders of magnitude.

> **b. The second category sets the standard that companies should provide <u>notice</u> to each user whose content is taken down or account is suspended**

---

[4] Heins, M. (2013). The brave new world of social media censorship. *Harv. L. Rev. F.*, *127*, 325.
[5] Cambridge Consultants. (2019). *Use of AI in Online Content Moderation*. https://www.ofcom.org.uk/__data/assets/pdf_file/0028/157249/cambridge-consultants-ai-content-moderation.pdf
[6] Hutchinson, B., Prabhakaran, V., Denton, E., Webster, K., Zhong, Y., & Denuyl, S. (2020). Social Biases in NLP Models as Barriers for Persons with Disabilities. *ArXiv:2005.00813 [Cs]*. http://arxiv.org/abs/2005.00813
[7] *Get Started with Perspective*. (n.d.). Perspective. Retrieved 30 June 2020, from https://www.perspectiveapi.com/#/start

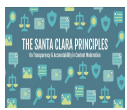

> **about the reason for the removal or suspension. Please indicate any specific recommendations or components of this category that should be revisited or expanded.**

The SCP call for platforms to notify users of how their "content was detected and removed (flagged by other users, governments, trusted flaggers, automated detection, or external legal or other complaint)." We believe that it is particularly important that users are notified of whether their content was flagged/removed by a software or by a human because this may impact how they perceive the credibility and reliability of the decision. Content-moderation rules are extremely rigid and are difficult to apply to the highly contextual variations of speech online. A *post-facto* content moderation method automatically applied by a software, is more likely to be perceived as unfair and unreliable by targeted users. This information may also influence their willingness to appeal the decision against them.

It is also important to find out the background of *why* exactly one's content was removed. In relation to this we would like to raise the following concerns. One issue is language: people might post in English, but they might make grammatical errors because English is not their mother tongue. In this case, it is possible that their post might be block-worthy, but not their intention behind it. Also, cultural variations should be taken into account: some things may be acceptable in one culture, but not in another. The worry is that standards and rules envisioned in the US may be hardly applicable globally and across the board: at least at the level of the appeal process, cultural and linguistic variations need to be taken into account.

As it relates to guidelines for content moderation, it will be key to involve as diverse and participatory an approach as possible in soliciting the most encompassing set of guidelines that are representative of the culture and norms of the geography within which the system is being deployed and used. Even within geographies, there is an argument to be made on deciding the granularity of the geographical region and reconciling that with the intra-region variations that might arise because of migrations of people with various cultures and languages from one part to another.

Finally, users should be notified of the measures that will be taken if more than one of their posts is removed and they should be aware of the consequences for posting content that violates the platform's standards. This is important because being banned from social media may have important psychological and social consequences, especially in countries in which social media play a key role in social interaction and information sharing.

> c. **The third category sets the standard that companies should provide a meaningful opportunity for timely appeal of any content removal or account suspension. Please indicate any specific recommendations or components of this category that should be revisited or expanded.**

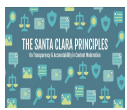

When it comes to content removal and account suspension, companies and organizations should be transparent about the timing and steps of the appeal. They should also make clear what kind of teams take care of the appeal process and how they deal with the culturally and linguistically specific circumstances mentioned above.

The process should also aim to be something that a lay individual can complete without having to seek external legal counsel as that itself might become a barrier for people to appeal decisions that they deem to be unfair if they don't have the means to access services that can help them appeal those decisions. Following a GDPR-like requirement for making the legalese accessible and understandable to the users, we propose that platforms also invest in handy tutorials, both text and video, that make it easier for people to complete the process on their own.

2. **Do you think the Santa Clara Principles should be expanded or amended to include specific recommendations for transparency around the use of automated tools and decision-making (including, for example, the context in which such tools are used, and the extent to which decisions are made with or without a human in the loop), in any of the following areas:**

    a. **Content moderation (the use of artificial intelligence to review content and accounts and determine whether to remove the content or accounts; processes used to conduct reviews when content is flagged by users or others)**

We strongly advocate for more transparency in content moderation. In particular, we think it is important that companies be as transparent as possible on the following issues:

**Standards** and rules implemented in content moderation: some platforms adopt broad standards that leave great room for discretion in their implementation. Other platforms adopt rules that are much more explicit, but are rigid and allow for less context variation. Either way, there must be an understandable explanation of the rules or standards moderators are operating with. One worry we have is that it is unclear what the ethical criteria are behind these provisions. In other words, what is the ultimate rationale for removing content or banning an account: is it because the post is offensive, abusive or harmful? An account is removed because it is inauthentic, because it disseminates falsehoods, or both? And how do companies define their moderation guidelines? These may differ widely depending on the platform, the organizational culture, the background, the business model, etc. In addition, more clarity is needed to guide users so they can better engage with online content.

**Agents** and actors behind content moderation: as mentioned earlier, when automatic decisions are taken, there has to be some indicator that there is an AI at work and whether or not humans have also been involved. And when humans are involved, it is important that we know about their role in the content-moderation process (are they supervising or being supervised?), their location/background and any training process they underwent to prepare them for their roles.[8] Also, to monitor content-curation, it is important to know whether these monitoring teams are internal or outsourced to other companies (which could potentially create a monopoly of content-moderating actors).

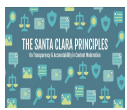

**Modality** and timing of content moderation: companies need to be explicit about the modality of their content moderation operations, namely whether they use mostly ex-ante vs. ex-post, reactive vs. proactive types of content moderation tools. They should also be open about the steps and timing of the moderating process.

We also advocate that this distinction above be quantified in the reporting provided by the platform as it will help users and the community better understand the extent to which content that is being moderated even makes it to a stage where it might be reviewed by a human, is reactive, compared to the amount of content that it being proactively being removed based on their established guidelines.

> b. **Content ranking and downranking (the use of artificial intelligence to promote certain content over others such as in search result rankings, and to downrank certain content such as misinformation or clickbait)**

We believe that content ranking could be an effective content moderation tool, but we also see potential risks:

**Transparency and bias** in content moderation: we believe that there is a general lack of transparency on how platforms guide content-ranking. When a post is downranked, the user will not notice or be notified, and this makes transparency an even more pressing concern; if there are insufficient measures to ensure moderators' biases do not interfere with their tasks there is more potential for the unfair downranking of posts, tailored to what the moderator believes. Importantly, the bias present in human moderators is bound to affect the data set being used to train the AI moderator too. This worry makes it urgent to **know who's doing the moderating and to figure out ways to make the data more representative**.

This can also quite easily morph into "shadow bans"[9] in which the platform can silently banish a user's content to algorithmic purgatory by pushing it down in rankings as a way of subversive censorship. Because of the lack of transparency on this, the practice needs to be addressed head-on in the principles.

---

[8] For instance, Facebook uses moderators located in the Global south and trains them to apply rules and provisions written in the US. The goal is to train moderators to override their initial judgements and reactions and apply Facebook guidelines.

[9] Forsey, C. (2020, October 30). Everything You Need to Know About Instagram's Secret Shadowban. *HubSpot*. https://blog.hubspot.com/marketing/instagram-shadowban

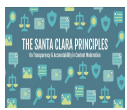

**Social change and freedom of expression**: freedom of speech is comparable to a spectrum that goes from hate speech to censorship (namely suppressing content that is deemed "unacceptable" for political reasons). When it comes to *online* content, freedom of expression is usually framed as the ability to post on online platforms, but we believe that it *also* relates to the ranking process on those platforms. Ranking of content is something that a platform does through its algorithm, and while it's not necessarily the responsibility of the moderator to ensure that users have amplified voices, the issue is whether in some cases it is their responsibility to downplay/downrank those voices. With respect to that issue, we are concerned that downplaying certain content may at times stifle the social progress that could be promoted by sharing sensitive content. A case in point here is the shooting of Philando Castile by a police officer in 2016. The shooting was recorded by Castile's girlfriend and streamed via Facebook Live. The video was initially taken down by Facebook because of its graphic content, but was later on reinstated.[10] The shocking and revealing content of the video could have a widespread impact and spark social change *only if* its ability to spread and be viewed repeatedly is preserved, even when it violates some of the platform's internal rules. Had the platform automatically downranked the video showing the shooting, we believe it would have limited freedom of expression and limited the possibility for widespread awareness on important social issues.[11]

Relatedly, it would be interesting to have insights from the platform on the degree of influence individual tastes have compared to collective averaging effects[12] that are used in surfacing and hiding content within the feeds of users. Specifically, if there are pervasive trends that are promoted by the underlying platform algorithms (say, towards higher profitability to the detriment of all other metrics) then that needs to be evident in the interest of transparency.

**Misinformation and users' autonomy**: In his "argument from truth", John Stuart Mill has argued that we should not restrict circulation of information, even when it is patently false. Indeed, we should permit perceived falsities to circulate and, with few exceptions, it is generally counterproductive to intervene in the free circulation of ideas by censoring some ideas and promoting others because this may stifle progress. Following Mill in *On Liberty*, then, one might argue that direct interventions to block or downplay misinformation online is called for *only* when it can be shown that it could cause harm to a group or to individuals. We worry that curating content for purposes other than to stop harm may limit users' autonomy in selecting the content they engage with. The notion of autonomy at play here comes from the work of philosopher Tom Scanlon.[13] This approach to free speech emphasizes the fact that we should be free to hear and have access to "what's said". Thus, we should be careful in limiting free expression just to protect consumers from forming false beliefs. It is preferable to let agents rely on their own reasoning powers in deciding what to believe. In this sense, direct interventions to block fake news sites, for instance, could be taking away consumers' autonomy as rational agents who are able to judge and make up their own minds. There are other solutions to limit the spread of misinformation online without impinging on consumers' freedom and autonomy. Behavioral nudges strategies are available and seem effective to fight fake news. There is evidence that if people are prompted to engage in careful reasoning, they become more alert about fake news. Indeed, these preventive measures seem to increase the likelihood consumers are less subject to accept misinformation.[14] Other behavioral nudges consist of pushing users to be mindful when sharing content that is potentially false or misleading.[15] The underlying premise backing up these solutions is that the spread of misinformation is the result of the fact that we under-employ our analytic reasoning and don't think reflectively when engaging with news online.[16] Accordingly, we can try to foster users' critical thinking skills when dealing with news online. As such, this approach is a form of nudging people into becoming better thinkers, but it is a form of nudging that *promotes* our autonomy and does not bypass our deliberative capacities. [17] As a result, this solution is preferable over other, more direct and invasive, ways of curbing fake news dissemination. **We believe that these intermediate solutions should be encouraged but we also see the need for clear guidelines to monitor when and how these behavioral tools are implemented.**

---

[10] Isaac, M., & Ember, S. (2016, July 8). Live Footage of Shootings Forces Facebook to Confront New Role. *The New York Times*.
https://www.nytimes.com/2016/07/09/technology/facebook-dallas-live-video-breaking-news.html

[11] See Garunay, M. (2016, July 7). *President Obama on the Fatal Shootings of Alton Sterling and Philando Castile*. The White House | President Barack Obama. https://perma.cc/VUL4-QT44.
We are nonetheless concerned about the potential voyeuristic aspect of letting representations of suffering and violence remain online. For instance, many White individuals have historically treated (and may continue to treat) representations of Black suffering (lynchings/slave auctions/shootings) as a form of entertainment for their viewing pleasure. (See Hartman, S. V. (1997). *Scenes of subjection: Terror, slavery, and self-making in nineteenth-century America*. Oxford University Press.) Allowing images or videos of violence and suffering could therefore perpetuate this harmful dynamic. In this sense, leaving sensitive content on social media could have adverse effects, especially on racialized individuals.

[12] Duportail, J., Kayser-Bril, N., Schacht, K., & Richard, É. (2020, June 15). *Undress or fail: Instagram's algorithm strong-arms users into showing skin*. AlgorithmWatch.
https://algorithmwatch.org/en/story/instagram-algorithm-nudity/

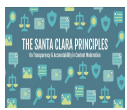

---

[13] Scanlon, T. (1972). A theory of freedom of expression. *Philosophy & Public Affairs*, 204-226.
[14] See Linden, S. van der, Leiserowitz, A., Rosenthal, S., & Maibach, E. (2017). Inoculating the Public against Misinformation about Climate Change. *Global Challenges*, 1(2), 1600008. https://doi.org/10.1002/gch2.201600008 and Cook, J., Lewandowsky, S., & Ecker, U. K. H. (2017). Neutralizing misinformation through inoculation: Exposing misleading argumentation techniques reduces their influence. *PLOS ONE*, *12*(5), e0175799. https://doi.org/10.1371/journal.pone.0175799.
[15] Porterfield, C. (2020, June 10). *Twitter Begins Asking Users To Actually Read Articles Before Sharing Them*. Forbes. https://www.forbes.com/sites/carlieporterfield/2020/06/10/twitter-begins-asking-users-to-actually-read-articles-before-sharing-them/#68ae71ee66a3
[16] Pennycook, G., & Rand, D. G. (2019). Lazy, not biased: Susceptibility to partisan fake news is better explained by lack of reasoning than by motivated reasoning. *Cognition*, *188*, 39–50. https://doi.org/10.1016/j.cognition.2018.06.011. In contrast, others have argued that fake news acceptance is the result of our partisan motivated reasoning. See Nyhan, B., & Reifler, J. (2010). When Corrections Fail: The Persistence of Political Misperceptions. *Political Behavior*, *32*(2), 303–330. https://doi.org/10.1007/s11109-010-9112-2 and Kahan, D. M. (2013). Ideology, motivated reasoning, and cognitive reflection. *Judgment and Decision Making*, *8*(4), 18.
[17] Levy, N. (2017). Nudges in a post-truth world. *Journal of Medical Ethics*, *43*(8), 495–500. https://doi.org/10.1136/medethics-2017-104153

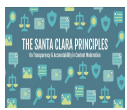

    c. **Ad targeting and delivery (the use of artificial intelligence to segment and target specific groups of users and deliver ads to them)**

When it comes to political advertising, we need to make sure that platforms are transparent on the integrity policies they are adopting and on whether these policies are the same as the policies adopted for other types of content. For example, although the public often does *not* differentiate between these two types of content, some platforms have adopted different internal policies to moderate "organic content" vs. political ads, creating the conditions for confusion and disinformation to spread. As Yaël Eisenstat explains, Facebook's "business model exploits our data to let advertisers aim at us, showing each of us a different version of the truth and manipulating us with hyper-customized ads." That means that we need transparency on the policies adopted across *all* sections of each online platform. What's more, when it comes to political advertising, it would be important that online platforms disclose what targeting tools have been used to tailor advertising campaigns on their sites.[18]

    d. **Content recommendations and auto-complete (the use of artificial intelligence to recommend content such as videos, posts, and keywords to users based on their user profiles and past behavior)**

Some of the same points as in previous answers apply here when it comes to how content recommendations are made. On the subject of auto-complete, there is a very subtle effect that will become pervasive which is around gradual nudging that redoes and potentially reduces humans' language use patterns and the diversity and richness of expressions of thought. Specifically, in the quick responses that are offered as options in the Gmail app on one's phone, the convenience of choice that lowers cognitive burden and provides crutches to start the response can steer how people reply to certain types of conversations, subtly chipping away at the agency of people to freely express themselves. This effect is also relatedly observed in smart voice assistants which nudge users to express themselves in particular ways, for example, phrasing and rephrasing requests and responses to the assistant till it conforms to what the machine understands. In complete opposition to the principle of having technology serve the needs of humans by adapting to their styles, humans will be forced to adapt their conversation styles to fit with that of the machine.

3. **Do you feel that the current Santa Clara Principles provide the correct framework for or could be applied to intermediate restrictions (such as age-gating, adding warnings to content, and adding qualifying information to content). If not, should we seek to include these categories in a revision of the principles or would a separate set of principles to cover these issues be better?**

---

[18] Eisenstat, Y. ( 2019, November 4). *I worked on political ads at Facebook. They profit by manipulating us.* Washington Post.
https://www.washingtonpost.com/outlook/2019/11/04/i-worked-political-ads-facebook-they-profit-by-manipulating-us/

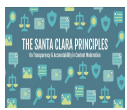

In general, we believe that **no one solution will fit all cases,** and so intermediate restrictions are called for. That means that content moderation tools will need to be tailored to specific issues: for instance, misinformation dissemination may be best addressed by using less stringent measures (e.g. behavioral nudges, as we explain above), whereas serious violations (e.g. hate speech) may require more drastic solutions such as content or account removals. There could be circumstances where intermediate restrictions could discourage reading a person's posts rather than just removing an individual post or user's account from the platform altogether. Because the SCP are primarily focused on violations and flagging/removal, areas in intermediate restrictions should follow a separate set of principles: we need clear standards to guide the moderating process in all cases, making sure it is fair and does not incur in the problems we highlighted above (e.g. bias).

4. **How have you used the Santa Clara Principles as an advocacy tool or resource in the past? In what ways? If you are comfortable with sharing, please include links to any resources or examples you may have.**

A participant in one of our public meetups discussing the SCP mentioned she had used the Principles as an advocacy tool. Based in Bangalore, India, she found that the Notice and Appeal Principles were most helpful in articulating how she believed a law regarding content moderation should be amended so as to not overcensor or stifle free speech online.

5. **How can the Santa Clara Principles be more useful in your advocacy around these issues going forward?**

In our work when we are asked about how to operationalize principles around content moderation, the SCP provides a succinct list of guidelines which are extremely helpful in orienting organizations that are just looking for practical guidance surrounding how they should go about the process of content moderation on their platforms in a way that adheres to some of the best practices and processes in the industry. We like the degree of concreteness offered by the SCP which allows for some flexibility in terms of implementation yet provides sufficient granularity to not be too abstract and vague which is something that we've seen in our work to be particularly challenging for organizations that want to try to adhere to responsible AI principles but don't have sufficient guidance on it.

Going forward, another recommendation that we feel will increase the impact of these principles is to explicitly ask agents in the ecosystem on what barriers they face in the adoption of principle sets like the SCP so that the guidance provided as a part of the principles or as addendums to the principles help the agents operationalize them in practice. This is the most important consideration in our experience as we have a lot of high-level guidance from a normative perspective but less so on tangible outcomes for engineering and design teams that are responsible for the actual development and deployment of these systems.

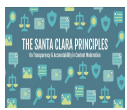

6. **Do you think that the Santa Clara Principles should apply to the moderation of advertisements, in addition to the moderation of unpaid user-generated content? If so, do you think that all or only some of them should apply?**

Yes, we strongly recommend that SCP apply to moderation of advertisements, please see above under question 2.c.

7. **Is there any part of the Santa Clara Principles which you find unclear or hard to understand?**

No, all parts are clear.

8. **Are there any specific risks to human rights which the Santa Clara Principles could better help mitigate by encouraging companies to provide specific additional types of data? (For example, is there a particular type of malicious flagging campaign which would not be visible in the data currently called for by the SCPs, but would be visible were the data to include an additional column.)**

Under question 1a, we pointed out that we are concerned about the **users' flagging/reporting system** as it could be particularly problematic when used in contexts where the majority of users are prone to discriminate against minority groups and thus see their posts as 'offensive'. As a result, we recommend that platforms disclose the numbers of initially *flagged* (by users) posts and how many of them are:
(i) <u>prioritized</u> by the platform and reported to the authorities because they raised serious concerns based on internal rules or standards;
(ii) categorized as <u>less urgent</u> (even if they are still violating internal rules or standards);
(iii) seen as *<u>not</u> violating* the platform's rules or standards (data on this could reveal patterns of discrimination against certain communities).

9. **Are there any regional, national, or cultural considerations that are not currently reflected in the Santa Clara Principles, but should be?**

What constitutes offensive speech may be a highly subjective, culture-specific issue and this may be a problem for platforms that operate on a global scale and use similar standards across the board. To moderate content on a global scale, we need to redefine our limits on what is acceptable and what isn't. We should have strong guardrails around what is acceptable behaviour so that basic rights of users are protected while being respectful of local cultural and contextual specifics.

10. **Are there considerations for small and medium enterprises that are not currently reflected in the Santa Clara Principles, but should be?**

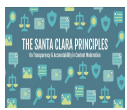

Small and medium enterprises (SMEs) face challenges that may at first be invisible when thinking about content moderation. Oftentimes, the first companies that come to mind are the classic tech giants -- Facebook, Twitter, WhatsApp, WeChat, Instagram, and others. Because of their size, SMEs have much less content posted on their platforms by users. For instance, an SME may only have to manage comments on an article (not pictures or videos), and only a small number of accounts relative to the world's most popular platforms. Nevertheless, it may not have the infrastructure or resources to adhere to the SCP, and appropriately collect and disclose the required information. While this is not an insurmountable barrier, the SCP should at least acknowledge the reality of SMEs and provide some guidance as to what measures they can take to abide by the principles.

We also advocate for the creation of public commons in terms of toolkits that help in adherence to the SCP, akin to the work done by organizations like the Tech Against Terrorism group[19] that has made a lot of tools that would typically be available to large corporations available to SMEs in an open-source and open-access manner allowing them to leverage the R&D investments made by larger actors in the ecosystem in preventing the promulgation of hate speech and other harms on their platforms.

## 11. What recommendations do you have to ensure that the Santa Clara Principles remain viable, feasible, and relevant in the long term?

We believe the Principles should be periodically revisited, for instance, every five years, or within a timeframe that allows for any appropriate revisions. This would allow the SCP to reflect various technological advancements, modifications in law and policy, as well as changing trends or movements in terms of platforms' content moderation. In addition, as explained in our answer to question 1b, it is crucial that the SCP reflect potential cultural and linguistic variations. We believe these considerations are also central to ensuring the SCP are viable, feasible, and relevant now and in the future. It may also be fruitful for those responsible for the maintenance of the SCP to consult with experts on how content moderation takes place on small, medium, and large companies' online platforms. While current practices in these areas may not reflect the ideals of the SCP yet, an in-depth understanding of these practices could greatly enrich the SCP. Understanding current practices can also help supplement the SCP with more precise examples and guidelines for those applying them. For example, this could take the broad form of "If you are a company of type x, implementing content moderation practices using resources z, then try to implement practices a and b". Such an approach could greatly increase the reach of the SCP by increasing their accessibility and ease-of-use. Ultimately lowering the barriers to adoption will lead to the SCP becoming more ubiquitous which is going to be essential to achieving the overarching goals of the SCP to build a safer, ethical, and more inclusive internet space for all of us.

---

[19] *Home*. (n.d.). Tech Against Terrorism. Retrieved 30 June 2020, from
https://www.techagainstterrorism.org/

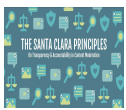

What is more, feedback should be sought from companies of different sizes known to have implemented the SCP, as well as from other companies who have not. This can help highlight what is most helpful within the SCP as well as what may be hindering adoption (for instance, lack of awareness, lack of resources, disagreement with the SCP, etc.) Creating at least some principles specific to human moderators and others specific to AI moderation could also contribute to the SCP remaining valuable and applicable to a large range of companies and their methods of moderation. Principles laying out the optimal process for a company looking to create or amend its content moderation policies and community guidelines could be of great help to companies. Principles regarding this should, among other things, address the importance of having diverse actors and perspectives at the table when creating the guidelines and policies that will regulate a platform. The SCP should likewise help bring to light the ethical and social implications of who gets to make guidelines and policies and how these are made and implemented to help push reflections beyond first-order impacts or considerations. Lastly, once a company has adopted the SCP, a mechanism should be put in place to not only hold them accountable but also to encourage continued adherence through time. If possible, we would also advocate the principles of transparency for the SCP itself in terms of the feedback that is received from various organizations that have tried them out, what they found useful, what hindered adoption, and other challenges that they might have faced which is going to be crucial in encouraging healthy debate around this and allowing for all stakeholders in the ecosystem to collaborate in developing them which will lead to a higher degree of ownership and hence adoption.

**12. Who would you recommend to take part in further consultation about the Santa Clara Principles? If possible, please share their names and email addresses.**

- Researchers doing work on disinformation/content moderation outside the Western context.
- Small, medium, and big companies that do content moderation.
- Individuals with experience either being content moderators or building content moderation AI.
- Partnership on AI: Media Integrity Team
- First Draft
- Witness Org.
- Montreal AI Ethics Institute [support@montrealethics.ai] -- We already have an ongoing project on misinformation related to climate change going on with other partner organizations that we believe would be relevant to this conversation around the SCP.

**13. If the Santa Clara Principles were to call for a disclosure about the training or cultural background of the content moderators employed by a platform, what would you want the platforms to say in that disclosure? (For example: Disclosing what percentage of the moderators had passed a language test for the language(s) they were moderating or disclosing that all moderators had gone through a specific type of training.)**

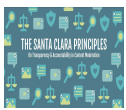

As we mentioned above, we believe that a disclosure about the training or cultural background of the content moderators employed by a platform, is highly recommended. We would need to know: the percentage of people undergoing certain training, their role in the content-moderating process (are they supervisors? Are they involved in the appeal process too?), their professional experience (e.g., are they lawyers?), their location and language, and whether these monitoring teams are internal or outsourced to other companies. It would also be useful to know what are the policies for their protection (e.g. physically, emotionally, mentally, etc.), what are the incentives that are offered to them, how is their performance measured, and other workplace conditions as this will shed light on to how the human content moderators are making their decisions.

**14. Do you have any additional suggestions?**

When it comes to messaging applications (e.g. WhatsApp), there is a tension between users' privacy protection and content moderation. Messaging apps encrypt messages to protect users' privacy and so for them it is hard to do any content moderation at scale. There have been proposals on how to go about doing some content moderation without breaking encryption[20] and we would like to see this addressed in the SCP.

**15. Have current events like COVID-19 increased your awareness of specific transparency and accountability needs, or of shortcomings of the Santa Clara Principles?**

The COVID-19 crisis has made clear for us that transparency surrounding "bots" -- automated accounts on social media platforms that often push misinformation or propaganda -- is lacking from companies. This is especially concerning in our current context, as bots are spreading conspiracy theories about the virus and false cures at an alarming rate[21], and as this misinformation could result in serious harm or death. Furthermore, the current context has also brought to light the international reality of content moderation; we should be looking to other countries and companies internationally and evaluating the advantages and drawbacks of their methods. The COVID-19 crisis has also made us reflect on the privilege we have in being able to discuss content moderation and misinformation publicly, as this is not everyone's reality. In light of this, we should also work towards preserving our ability to have these conversations openly.

It was also made clear very recently that content moderation can get extremely political, perhaps even more than we had previously realized, as tweets by President Donald Trump were marked by Twitter content moderators as partially false/misleading), and encouraging users to "get the facts".[22] This event, and the implications surrounding taking action (or refusing to) against a public figure like President Trump by moderating his tweets seem to warrant more guidance on how public figures should be treated. At the least, a discussion centered on what is at stake when public figures are subjected or not to content moderation appears necessary. The implications we see are the following: the important differences in impact if a post is marked as misleading or false vs. if it is fully removed, risks relating to limiting free speech, risks concerning an oversized amplification of falsehoods, concerns about holding all users to the same standards, and concerns relating to people having an accurate picture of what a public figure (like a President) believes and puts forward.

---

[20] Reis, J.C., Melo, P.D., Garimella, V.R., & Benevenuto, F. (2020). Detecting Misinformation on WhatsApp without Breaking Encryption. *ArXiv, abs/2006.02471*.

[21] Pesce, N. L. (2020, May 26). *About half of the Twitter accounts calling for reopening America are bots: Report*. MarketWatch.
https://www.marketwatch.com/story/about-half-of-the-twitter-accounts-calling-for-reopening-america-are-probably-bots-report-2020-05-26

---

[22] Wong & Levine (2020, May 26). Twitter labels Trump's false claims with warning for first time. *The Guardian*. https://www.theguardian.com/us-news/2020/may/26/trump-twitter-fact-check-warning-label

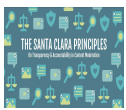